\documentclass[fleqn,twoside]{article}
\usepackage{espcrc2}
\usepackage{graphicx}
\usepackage[figuresright]{rotating}

\newcommand{\AmS}{{\protect\the\textfont2
  A\kern-.1667em\lower.5ex\hbox{M}\kern-.125emS}}

\title{A triple-GEM telescope for the TOTEM experiment}

\author{S. Lami\address[INFN]{Pisa INFN, Largo B. Pontecorvo, 3 - 56127 Pisa, Italy},
        G. Latino\address[SIENA]{Physics Department, Siena University,
	 Via Roma, 56 - 53100 Siena, Italy}\addressmark[INFN]%
	 \thanks{Corresponding author. Phone: +39-050-2214439. E-mail address: giuseppe.latino@pi.infn.it}, 
	E. Oliveri\addressmark[SIENA]\addressmark[INFN],
	L. Ropelewski\address[CERN]{CERN, EP Division, 1211 Geneva 23, Switzerland},
	N. Turini\addressmark[SIENA]\addressmark[INFN]}
       
\begin{document}

\begin{abstract}
The TOTEM experiment at LHC has chosen the triple Gas Electron
Multiplier (GEM) technology for its T2 telescope which will provide 
charged track reconstruction in the rapidity range 5.3$<$$|\eta|$$<$6.5
and a fully inclusive trigger for diffractive events.
GEMs are gas-filled detectors that have the advantageous decoupling of 
the charge amplification structure from the charge collection and readout 
structure. Furthermore, they combine good spatial resolution with very 
high rate capability and a good resistance to radiation.
Results from a detailed T2 GEM simulation and from laboratory tests on a 
final design detector performed at CERN are presented.
\vspace{1pc}
\end{abstract}

\maketitle
\section{INTRODUCTION}
\begin{figure}[htb!]
\vglue -0.3in
\hglue 0.2in 
\includegraphics[scale=0.3]{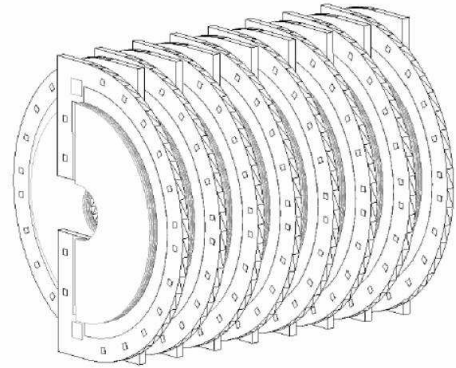}
\vglue -0.3in
\caption{One arm of TOTEM T2 Telescope.}
\label{fig:T2_telescope}
\vglue -0.3in
\end{figure}

The TOTEM~\cite{Totem} experiment at the LHC collider will measure the total $pp$ 
cross section with a precision of about 1$\div$2\,$\%$, the elastic $pp$ cross section 
over a wide range in -t, up to $10$\,GeV$^2$, and will study diffractive dissociation 
processes.
Relying on the ``luminosity independent method'' the evaluation of the total cross section 
with such a small error will require simultaneous measurements of the $pp$ elastic  
scattering cross section $d\sigma /dt$ down to $-t \sim 10^{-3}$\,GeV$^2$ (to be extrapolated to 
$t$ = 0) as well as of the $pp$ inelastic interaction rate with a good rapidity 
coverage up to the very forward region. 
Roman Pots (RP) stations at 147\,m and at 220\,m on both sides from the Interaction Point (IP), 
equipped with ``edgeless planar silicon'' detectors, will provide the 
former measurement. The latter will be achieved by two inelastic telescopes, T1 and T2, 
placed in the forward region of the CMS experiment on both sides of the IP. 
T1, using ``Cathode Strip Chambers'', will cover the rapidity range 3.1$<$$|\eta|$$<$4.7 
while T2, based on ``Triple-GEM'' technology, will extend charged track reconstruction 
to the rapidity range 5.3$<$$|\eta|$$<$6.5.
These detectors will also allow common CMS/TOTEM diffractive studies  
with an unprecedented coverage in rapidity.
The T2 telescope will be placed 13.56\,m away from IP and 
the GEMs employed will have an almost semicircular shape, with an inner radius 
matching the beam pipe. Each arm of T2 will have a set of 20 triple-GEM detectors 
combined into 10 aligned semi-planes mounted on each side of the vacuum pipe 
(Figure~\ref{fig:T2_telescope}).
\section{GEM TECHNOLOGY}
The CERN developed GEM technology~\cite{GEM_Sauli} has already been successfully 
adopted in other experiments such as COMPASS and LHCb and has been considered for 
the design of TOTEM very forward T2 telescopes thanks to its characteristics: large 
active areas, good position and timing resolution, excellent rate capability and 
radiation hardness.  
Furthermore, GEM detectors are also characterized by the advantageous decoupling of 
the charge amplification structure from the charge collection and readout 
structure which allows an easy implementation of the design for a given apparatus.  
The T2 GEMs use the same baseline design as the one adopted in COMPASS~\cite{GEM_Compass}: 
each GEM foil consists of thin copper clad polymer foil of 50\,$\mu$m, with copper layers of 
5\,$\mu$m on both sides, chemically perforated with a large number of holes of
70\,$\mu$m in diameter. A potential difference around 500\,V applied between the two 
copper electrodes generates an electric field of about 100\,kV/cm in the holes which 
therefore can act as multiplication channels (gains of order $10 \div 10^2$) for electrons 
created in a gas (Ar/CO$_2$ (70/30 $\%$) for T2) by an ionizing particle. 
The triple-GEM structure, realized by separating three foils by thin (2$\div$3\,mm) insulator 
spacers, is adopted in order to reduce sparking probabilities while reaching high total gas 
gains, of order $10^4 \div 10^5$, in safe conditions. 
The read-out boards will have two separate layers with different patterns: one with 256x2 concentric
circular strips, 80\,$\mu$m wide and with a pitch of 400\,$\mu$m, allowing track radial reconstruction 
with ${\sigma}_R$ down to 70\,$\mu$m, and the
other with a matrix of 24x65 pads of 2x2 to 7x7\,mm$^2$ in size from inner to outer circle, 
providing level-1 trigger information as well as track azimuthal reconstruction.
\section{T2 TRIPLE-GEM SIMULATION} 
\begin{figure}[htb!]
\vglue -0.3in
\includegraphics[width=1.0\linewidth]{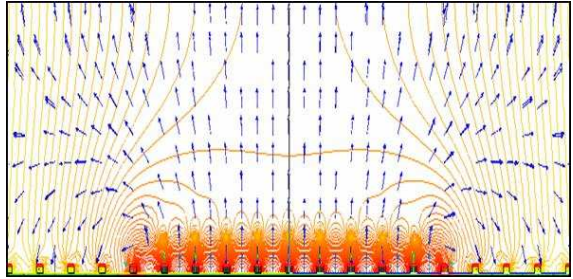}
\vglue -0.2in
\caption{Simulation of the weighting field for a T2 GEM pad electrode.}
\label{fig:Pad_Weighting_Field}
\vglue -0.4in
\end{figure}

A detailed simulation of T2 triple-GEM detector has been developed starting from the 
existing implementation for the GEMs used at LHCb~\cite{GEM_Sim}. The general framework is 
relying on several packages allowing a complete and detailed ``step by step'' simulation, 
for a given gas mixture and detector geometry, for the several underlying processes: 
starting from the primary ionization up to the spatial and timing properties of the collected signals. 
The main framework is implemented in {\em Garfield}; the electric field mapping is simulated 
with {\em Maxwell}; the electron/ion drift velocity and diffusion coefficients are evaluated 
with {\em Magboltz}; Townsend and attachment coefficients are simulated by {\em Imonte}; 
the energy loss by a given ionizing particle in gas and the cluster production process are 
evaluated by {\em Heed}.  
As an example, Figure~\ref{fig:Pad_Weighting_Field} reports the simulation for the ``weighting 
field'' ${\vec{E}}_K^W(x)$ (defined by putting at 1\,V the given readout electrode while 
keeping all the others at 0\,V) for a pad electrode. 
Signal induction is then derived via the Ramo theorem: 
$I_k = -q\vec{v}(x)\times {\vec{E}}_K^W(x)$. 

From the reconstruction of the full process chain leading to signal collection, 
with proper modeling of lateral electron cluster diffusion through each GEM foil, 
the expected signal for a MIP particle has been derived for both strips and pads for 
typical values of the electric field in the drift and induction zones between GEM foils (E$_{d/t}$ 
$\sim$ 3kV/cm). Timing properties, such as a typical signal time delay(duration) of $\sim$\,60(50) ns, 
have been found consistent with preliminary test beam studies on prototypes. 
Furthermore, the study of signals as a function of distance from electron cluster centroid, 
when combined with expected signal processing by the readout electronics, has shown 
a typical strip cluster size of 2$\div$3 channels (1$\div$2 for pads), which is consistent 
with COMPASS test beam results~\cite{GEM_Compass}. 
Ongoing test beam activities, performed with final production 
detectors read by final design electronics (digital readout via VFAT chip), are expected to allow 
an improved test and tuning of current simulation. 
\section{TEST ACTIVITIES AT CERN}
\begin{figure}[htb!]
\vglue -0.3in
\includegraphics[width=1.\linewidth]{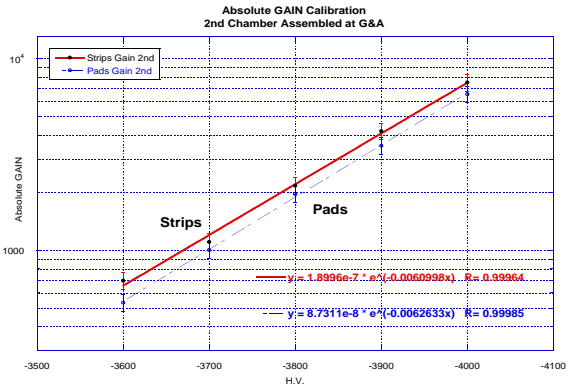}
\vglue -0.4in
\caption{Strip/Pad gain as a function of the applied high voltage.}
\label{fig:GainCalib}
\vglue -0.35in
\end{figure}

Two final full size detectors, whose components were provided by CERN, have been assembled 
by an italian private company~\cite{GeA}, 
and then tested at CERN Gas Detector Development Laboratory with a Cu X-Ray source 
($K_{\alpha /\beta}$ = 8/8.9 KeV). These activities involved studies on: 
general functionality, absolute gain, strip/pad charge sharing, energy resolution, time 
stability and response uniformity. In particular, the analysis of signals simultaneously collected 
from 8 strip/pad electrodes allowed to check the most important detector parameters. 
Figure~\ref{fig:GainCalib} shows the total effective gain $G_T$
\footnote{
This parameter can be derived according to equation
$G_T = I_{tot}/(e\cdot n\cdot f)$, 
from total readout current ($I_{tot}$) and X-Ray interaction rate ($f$) measurements, knowing the 
average number of electrons produced by an interacting X-Ray ($n$).
} 
for both strip and pad readout channels as a function of the applied HV: 
an expected gain of ~8$\div$10$\times$$10^3$ for a typical HV value of -4\,kV is observed. 

\begin{figure}[htb!]
\vglue -0.3in
\hglue -0.05in
\includegraphics[angle=-90,width=1.0\linewidth]{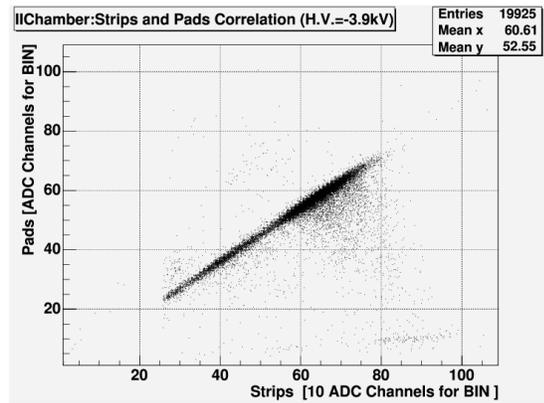}
\vglue -0.3in
\caption{Strip/Pad cluster charge sharing. The reduced correlation at higher current values 
(around the 8\,KeV peak) is attributed to not complete alignment of X-Ray beam to the instrumented 
readout channels. 
}
\label{fig:ChShar}
\vglue -0.3in
\end{figure}

The study of strip/pad cluster charge sharing showed the expected correlation between the 
two clusters (Figure~\ref{fig:ChShar}). A slightly higher charge collected 
by strips (about 10$\div$15$\%$), considering the typical higher strip cluster size, 
is consistent with the design for an optimal setup of the common readout chip.   

The evaluation of energy resolution represents another important detector 
test as it is related to the quality and uniformity of GEM foils. In fact, 
a not uniform gain over the irradiated zone will results in an anomalous 
broadening of the peak in the response spectrum. An energy resolution of 
$\sim$ 20$\%$, in terms of FWHM for the leading 8 KeV peak, was found to be well 
in agreement with the expected design performance of the detector.

Furthermore, time stability of signal has been tested with continuous detector 
irradiation over more than one hour and response uniformity checked by 
randomly moving the X-Ray source over the detector surface. 

In conclusion detector performances well within 
expectations have been observed. A more extensive test on ten production detectors 
will be performed at the incoming test beam activities.  
\vspace{-0.5pc}
\section*{ACKNOWLEDGMENTS}
\vspace{-0.5pc}
We are particularly grateful to A. Cardini and D. Pinci of LHCb Collaboration 
for very precious initial input on triple-GEM simulation.    

\end{document}